\documentstyle[prb,aps,multicol,psfig]{revtex}
\tighten
\begin{document}
\preprint{\today}
\title{Small polaron formation in the Holstein and Su-Schrieffer-Heeger
models: The criteria from analytic and numerical analyses}
\author{M. Capone, M. Grilli and W. Stephan\cite{newaddress}}
\address{Istituto Nazionale di Fisica della Materia and
Dipartimento di Fisica, Universit\`a di Roma ``La Sapienza'',\\
Piazzale A. Moro 2, Roma, Italy 00185}
\maketitle

\begin{abstract}
We investigate the conditions leading to polaron formation
for a single electron interacting with dispersionless 
optical phonons within the Holstein and the Su-Schrieffer-Heeger
models. Both from analytic perturbation theory and 
exact numerical diagonalization of small clusters, we find different
criteria for the model parameters ruling the setting  in of the 
polaronic regime. We also illustrate the common physical origin 
of polarons in the two models as arising from the simultaneous
fulfilling of two conditions:  A sizable ionic displacement
and a lattice deformation energy gain larger that the loss
in the bare electron kinetic energy.
\end{abstract}


\begin{multicols}{2}

\section{Introduction}
Polarons have been observed in various materials
like ${\rm Ti_4 O_7}$, ${\rm Ni_x V_2O_5}$ and tungsten
oxides ${\rm W O_{3-x}}$. More recently polaronic features
have been detected in optical spectroscopy in the lightly
doped insulating phase of cuprate superconductors 
\cite{polaronexper,polaronexper2}.

Electrons acquire a polaronic character
in the presence of a sufficiently strong electron-phonon
($e$-$ph$) coupling when they displace the ions
around them and move carrying along the lattice
deformation. Being accompanied by the much heavier
lattice degrees of freedom, in the polaronic regime 
the carriers acquire large effective masses and, in some
cases, may even be trapped in the potential well arising
from the ionic displacement that they created.

Polarons are usually classified as being large or
small polarons depending on whether the ionic deformation 
is spread  over several sites or involves one single site.
In this paper we confine ourselves to the investigation of  
models with short-range $e$-$ph$ interaction, where, apart 
from a more or less narrow crossover region only small 
polarons are formed \cite{1dlp}.

In particular we investigate the Holstein model\cite{holstein} 
and the Su-Schrieffer-Heeger (SSH) model\cite{ssh}.
Due to their relative simplicity  these
models are definitely the most frequently
considered models for electrons and phonons
interacting via a short-range potential and
may well be taken as a suitable paradigmatic
basis for investigating the physics of strongly
interacting $e$-$ph$ systems. 

The concepts underlying polaron theory in these models
are long-standing and have found various theoretical
substantiations over the last decades. However, 
the strong-coupling nature of the polaronic state does 
not allow reliable analytic approaches in the intermediate
crossover region, where one should sit in order to
quantitatively investigate the conditions for polaron formation.
On the other hand the multiphononic essence of polarons
makes it difficult to approach the
strong-coupling regime from the numeric point of view.
A landmark in this context was provided by quantum
Monte Carlo calculations in Ref.\onlinecite{deraedt1a}
where an interpolation formula was presented
describing the critical $e$-$ph$ coupling leading
to polaron formation in the Holstein model. However, 
no detailed analysis and distinction between the 
adiabatic and anti-adiabatic regimes was carried out 
there and the interpolation formula
was presented in a rather empirical way. More recently
the discovery of polarons in the insulating phases 
of the high temperature superconductors has triggered
numerical exact diagonalization analyses on models with
strong $e$-$e$ interactions \cite{rt,wellein,greco}.

Despite the renewed interest, the literature still 
lacks a unifying picture providing a clear
physical and formal understanding of the phenomenon of
polaron formation both for the simple dilute case of one
single polaron in an empty lattice and for the case 
of many interacting electrons. The main goal of this
paper is to investigate
the Holstein model and the SSH model in order to provide
such a picture for the single particle case.

In the Holstein model
\begin{equation}
\label{holham}
{\cal H}=-t\sum_{\langle ij\rangle}
c^\dagger_{i}c_{j} +
g\sum_{i} 
c^\dagger_{i}c_{i} \left( a_i+a^\dagger_i \right)
+\omega_0 \sum_i a^\dagger_i a_i
\end{equation}
and in the SSH model
\begin{eqnarray}
{\cal H} & = & -t\sum_{\langle ij\rangle}
c^\dagger_{i}c_{j} 
+\omega_0 \sum_i a^\dagger_i a_i \nonumber \\
+ & g & \sum_i 
\left[ \left( c^\dagger_{i}c_{i+1} +
 c^\dagger_{i+1}c_{i} \right)
\left( a^\dagger_{i+1}+a_{i+1}-a^\dagger_i-a_i \right) \right]
\end{eqnarray}
the first term is proportional to 
the nearest-neighbor hopping 
integral $t$, which we will take as our unit 
of energy, giving  rise to a d-dimensional
tight-binding band structure of the form
$E({\bf k})=-2t \sum_{\nu =1}^d\cos(k_\nu)$
(for simplicity we assume a cubic lattice with unit
lattice spacing and $\hbar=c=1$) accounting for the 
kinetic energy of the free electrons.
Since we will restrict ourselves to the single electron
case we will not consider electron spin indices throughout
this paper.

A dispersionless ($\omega(q)=\omega_0$) Einstein
phonon is created by the field $a^\dagger_i$ and is coupled
to the local electronic density in the Holstein model
and to the covalent bond variable $ c^\dagger_{i}c_{i+1}+
c^{\dagger}_{i+1}c_i$ in the SSH model.  
For the Holstein model, the coupling  arises from the 
dependence of the local atomic energy (i.e. the Madelung energy)
on the ionic position.  This coupling is relevant when the
screening of the Madelung potential is poor, and
is believed to be non-negligible in the superconducting 
cuprates\cite{zeyher}. On the other hand the covalent
$e$-$ph$ coupling in the SSH model is due to the
dependence of the hopping integral on the relative
distance between two adjacent ions\cite{ssh}. 
Notice that our SSH model differs from the conventional one
in having optical (instead of acoustic) phonons like the
Holstein model: To clarify more easily the common mechanisms underlying
the polaron formation in the two models we avoided unnecessary 
differences between them, thus focusing on the role played
by the different $e$-$ph$ couplings. Moreover, despite the 
completely different origin of the $e$-$ph$ coupling
in the two models, we choose the same notation $g$ to
emphasize the generic character of the  physical 
processes that we are going to present.  

Before addressing the problem of the single polaron formation
in a more formal way within the above models, we first would
like to provide simple and intuitive arguments. As mentioned
above the setting in of a polaronic regime is characterized by
{\it both} a lattice deformation energy gain larger than the loss
of bare kinetic energy
{\it and} a sizable local displacement of the ionic
positions, giving rise to a strong reduction of the 
effective hopping matrix element.  

For the Holstein model, these effects are directly related to two
 parameters which are often introduced in this field:
$\lambda \equiv g^2/(2dt\omega_0)$ and 
$\alpha \equiv g/\omega_0$. The former has
a twofold meaning. In fact it can represent
the effective phonon-mediated $e$-$e$ coupling
introduced, e.g. in Fermi liquid theory or in the
traditional BCS theory of superconductivity.  Alternatively
and more interestingly in the present context, $\lambda$
may represent the ratio between the polaronic
binding energy $E_p=-g^2/\omega_0$ (see below)
in the strong coupling limit and the {\it bare} average
kinetic energy of the electrons  of the order
of half the bandwidth $(\sim -2td)$. Notice that the bare 
hopping $t$ has to be used here. In fact this is of the 
order of the kinetic energy actually lost when the polaron
is formed. Then the value of $\lambda$ determines the convenience
for the system to give up the kinetic energy gain
arising from the hopping to gain the lattice deformation
energy induced by the local $e$-$ph$ potential.
On the other hand, as is clear  from a
standard Lang-Firsov transformation of
the Holstein model\cite{langfirsov},
$\alpha$ represents half
the ionic displacement in units of $(2M\omega_0)^{- {1 \over 2}}$,
where $M$ is the ionic mass. 
{\it The simultaneous
occurrence of the two conditions $\lambda >1$ and
$\alpha >1$ is needed to characterize and to determine the
polaron formation in the
Holstein model}. Thus one can immediately recognize from
the definition of $\lambda$ and $\alpha$ that a crucial
role is played by the adiabatic ratio $\omega_0/t$.
If  $\omega_0/t$ is small, the condition  for a large
$\lambda=\alpha^2 \omega_0/2td$ is more difficult to
realize than $\alpha >1$, and polaron formation
will be determined by the more restrictive $\lambda >1$ 
condition. The opposite is true when the system is in the
anti-adiabatic regime  $\omega_0/t>1$. 
This intuitive argument was already implicit in the 
interpolation formula Eq. (4.1)
in Ref. \onlinecite{deraedt1a} once this is expanded in the
two opposite limits   $\omega_0/t \gg 1$ and $\omega_0/t \ll 1$.

For the SSH model it will be shown that the value of
$\lambda$ determines {\it both} the the tendency towards 
the localization of the electron {\it and} the suppression
of the hopping integral associated with the lattice distortion,
regardless the value of the adiabatic ratio $\omega_0/t$.
Therefore $\lambda$ will be the relevant parameter for
the description of the system for any value of $\omega_0/t$. 
This intuitive argument will be substantiated below
by analytic calculations and numerical exact diagonalization
of small clusters.

\section{A first insight from perturbative calculations}
The above  arguments may be made more formally precise
within a perturbative analytic calculation 
in the limit of small $e$-$ph$ coupling ($g \ll t,\omega_0$). 
In this case we evaluated the second order correction
to the electronic self-energy represented in the
diagram of Fig. 1. 

\begin{figure}
\vspace{-1.5truecm}
{\hbox{\psfig{figure=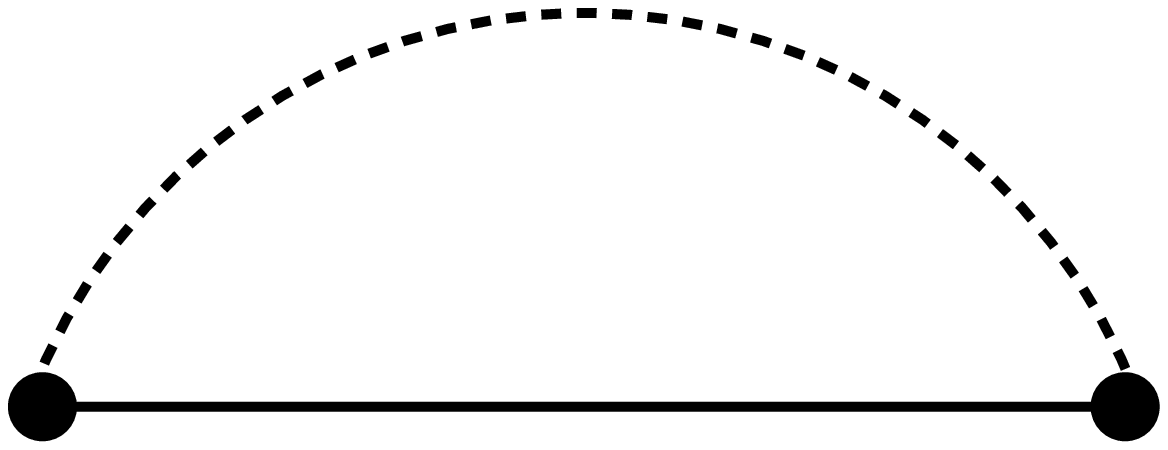,width=5.5cm,angle=-90}}}
\label{selfen}
\end{figure}
\vspace{-2truecm}
{\small FIG. 1: Lowest-order self-energy diagram
of the electron propagator. The solid line is the bare
electron Green function; The dashed line is the bare phonon propagator.
The dots represent the $e$-$ph$ coupling.}
\vspace{0.5truecm}

To explore both the adiabatic and the
anti-adiabatic regimes, we cannot apply Migdal's
theorem to discard vertex corrections: Our approximation
only  relies on the small value of $g$. This also 
allows for the simplification of using the bare electronic
Green function  instead of
carrying out a self-consistent evaluation including the
full Green function inside the self-energy diagram.

For the sake of simplicity and to allow for a more direct 
comparison with the numerical results on small clusters,
we only deal with one-dimensional systems. Nevertheless,
we explicitly checked in the easier case of the Holstein
model that our analytic results apply to multidimensional
cases as well. 

For our purposes we only need
 the perturbative corrections to the  effective mass 
\begin{equation}
{m^* \over m}={1- {\partial Re\Sigma(k,\omega)
\over\partial\omega}\vert_{\omega=-2t,\,k=0}
\over 
1+{\partial Re\Sigma(k,\omega)
\over \partial E_k}\vert_{\omega=-2t,\,k=0}}
\label{effmass}
\end{equation}
and to the ground state given by the solution of the equation 
\begin{equation}
\omega-\epsilon_0-Re\Sigma(k=0,\omega)=0.
\label{groundstate}
\end{equation} 
It is quite obvious that the polaronic regime cannot
be attained within our lowest-order
perturbative approach: The cloud dressing the electrons 
in a  polaronic excitation involves multiphononic processes,
which are not included in the diagram of Fig. 1.
Nevertheless valuable indications on the beginning
of the polaronic crossover can be extracted from the
above expressions. In particular, one can determine the
parameters for which the effective mass starts to 
grow [$(m^*-m)/m \sim 1$], also marking the
region where perturbation theory is no longer applicable.

\subsection{The Holstein model}
For the Holstein model, in agreement with Ref.\onlinecite{marsiglio} 
we obtained
\begin{equation}
\Sigma (\omega) ={ \lambda \omega_0 \over
\sqrt{ \left( {\omega - \omega_0 \over 2t} \right)^2 - 1}}
\end{equation}
where the real part of the square root has the same sign
as $(\omega-\omega_0)$. Notice that the self-energy
is momentum independent because the bare $e$-$ph$
vertex is also momentum independent. This feature
allows the introduction of the density of states in
the momentum integrals, thus leading to a straightforward
extension of our results above one dimension.

In $d=1$ we find
\begin{eqnarray}
{m^* \over m}&=&1+{2\lambda t \left(2t+\omega_0 \right) \over
\sqrt{\omega_0}\left(4t+\omega_0 \right)^{3/2}}
\label{massh} \\
E(0) & =& -2t \left( 1+ \lambda \sqrt{{\omega_0 \over \omega_0+4t}}
\right).
\label{groundh}
\end{eqnarray}
By expanding the perturbative correction in Eq.(\ref{massh})
in the two opposite, adiabatic 
($\omega_0 \ll t$) and anti-adiabatic\cite{notaantiad}
($\omega_0 \gg t$) limits, we get
\begin{eqnarray}
m^* & = & m\left(1+{\lambda \over 2}
\sqrt{t\over \omega_0} \right) \;\;\; \omega_0 \ll t \label{mhad} \\
m^* & = & m\left(1+\alpha^2 \right) \;\;\;\;\;\;\;\;\; \omega_0 \gg
 t \label{mhaad}
\end{eqnarray}
In agreement with  the intuitive arguments presented at the end
of the previous section, 
the anti-adiabatic result shows that the mass enhancement
is driven by the condition $\alpha >1$, which is more
restrictive than $\lambda >1$. This finding also
matches well with the result one would obtain from a
standard  Lang-Firsov transformation\cite{langfirsov} 
of the Holstein model. As is well known,
this transformation changes the free electron basis
into a polaron basis and shifts the ionic equilibrium
position by $\alpha$. By suitably choosing 
$\alpha=g/\omega_0$  the  coupling between
the phonons and the local electron density
is eliminated and the fermion-phonon
interaction is instead displaced into the kinetic energy term.
In the anti-adiabatic limit, where $t$ is small, this scheme
is more convenient (the interaction is put in the small
part of the Hamiltonian) and, averaging the Hamiltonian on
the phononic vacuum one finds a bare polaron binding
energy $E_p=-\alpha^2 \omega_0$ and an exponential reduction
of the kinetic energy $t \to t^*=t\exp(-\alpha^2)$. This standard
result can easily be connected to our perturbative result
(\ref{mhaad}) by noticing that, for small $g$, $\alpha$ is
also small so that $t^* \approx t(1-\alpha^2)$.

On the other hand, as expected, the adiabatic result for the 
mass enhancement is proportional to $\lambda$.
However, this contribution is also proportional to
$\sqrt{t/\omega_0}$, which is large in this limit
and this also seems to contrast the usual perturbative
calculations within Fermi liquid theory\cite{agd},
which only predict a correction of order $\lambda$.
This feature arises from the singular density
of states at the bottom of a one-dimensional band,
as can be  checked by considering
a finite density of electrons with a finite
Fermi energy $\mu$ away from the bottom of the 
band. In this latter case the coefficient of the mass
correction becomes $\sqrt{t/(\omega_0+\mu)}$
and is no longer singular in the adiabatic limit\cite{cappelluti}.
This specific, non-generic result is
the price that we have to pay in order to
take advantage of the simpler analytic treatment 
in one dimension, but it does not hide the
important finding that only the $\lambda$
parameter rules the polaron formation when
$\omega_0 < t$. Moreover it is worth noting
that our simple, lowest-order perturbative calculation 
gives a strong indication that one electron in a fully 
adiabatic\cite{notaadiab} one-dimensional Holstein lattice
is localized ($m^* \to \infty$). In light of
our calculation, this  well known result\cite{kabanov}
can easily  be attributed to the singular density of 
states\cite{perturb}.

\subsection{The SSH model}

The simplifying feature of a momentum independent
 bare $e$-$ph$ vertex is no longer present 
in the SSH model, where the bare $e$-$ph$ vertex in $d=1$
associated with scattering of an electron from a $k$ to a
$k+q$ Bloch state has the following form
\begin{equation}
g_{k,k+q}=2ig \left[ \sin(k+q) -\sin(k) \right].
\label{barevertex}
\end{equation}
Consequently, once a frequency integration is carried out,
the self-energy is given by
\begin{eqnarray}
\Sigma (k,\omega) & = & {4g^2 /N} \sum_q
{ 1 \over \omega - \omega_0 -\epsilon_{k+q} +i \delta}\nonumber \\
&\times & \left[\sin^2(k+q) +\sin^2(k)-2\sin(k)\sin(k+q) \right].
\label{selfenintegr}
\end{eqnarray}
 To obtain the desired physical quantities for a
single electron in the bottom of the band,
we only need to evaluate $Re\Sigma(k=0,\omega)$
and its derivatives in the $k=0$ state and for 
$\omega=E(k=0)=-2t$, finding
\begin{equation}
{m^*\over m}  =  1+ \lambda
\left[
{8\omega_0 \over 
\sqrt{\omega_0^2+4t\omega_0}} +
{2 \omega_0 \over t} \left( {2t+\omega_0 \over
\sqrt{\omega_0^2+4t\omega_0}} -1 \right) \right] 
\label{massssh} 
\end{equation}
\begin{equation}
E(0)  =  -2t \left(  1 + 2\lambda {\omega_0 \over t}
+\lambda {\omega_0^2 \over t^2}
\right). \label{groundssh}
\end{equation}
Also in this case we evaluate the  mass
correction in the adiabatic and in the anti-adiabatic limits
\begin{eqnarray}
m^* & = & m\left(1+6\lambda 
\sqrt{ \omega_0\over t} \right) \;\;\; \omega_0 \ll t \label{msshad} \\
m^* & = & m\left(1+8\lambda+2\alpha^2 \right) \;\;\; \omega_0 \gg
 t \label{msshaad}
\end{eqnarray}
Contrary to the Holstein model case,  the electrons in
the SSH model in the fully adiabatic limit
are completely free ($m^*=m$).
This is due to the vanishing of the bare $e$-$ph$ vertex
for small transferred momenta $q$ [cf. Eq.(\ref{barevertex})],
which overcompensates the divergent density of states 
in the integral of Eq. (\ref{selfenintegr}). Physically
this effect arises because many electronic states
lie close to the $k=0$ point, but the phonon-mediated
scattering between them occurs at low momentum transfer,
which is less effective in the SSH model, 
where phonons decouple from the electrons
in the long-wavelength limit. 

An important difference between the Holstein and the SSH
models is also present in the anti-adiabatic limit, where
the mass enhancement in the SSH model [Eq.(\ref{msshaad})]
involves both $\alpha^2$ and $\lambda$. In this limit
$\lambda \gg \alpha^2$ and therefore the mass
correction is dominated by $\lambda$ also in the
anti-adiabatic regime. This difference arises from the
different localization mechanism taking place in the SSH model
with respect to the Holstein model.
In light of this specific mechanism one can interpret the 
dependence of the mass enhancement on the parameter
$\lambda$ in both the adiabatic and anti-adiabatic
regimes. However, we prefer to postpone the discussion of this issue
until the full numerical analysis of the SSH model is
presented in the second part of the next section.
This analysis will provide a clear substantiation
of the above perturbative indications and a 
natural interpretation will be given of the criterion 
for polaron formation.

\section{Exact diagonalization analysis}

In order to found the above scheme for 
the single polaron formation on more solid ground,
we performed exact numerical calculations on
small clusters by means of the Lanczos algorithm.
As usual\cite{marsiglio}, we truncate the phononic
Hilbert space so as to include only a finite number
of phonons per lattice site. To reliably
 explore the strong-coupling regimes, we had 
to include up to 50 phonons per site (and check the
convergence of the results by varying the phonon
number). Due to the huge enlargement of the
Hilbert space induced by the presence of the lattice
degrees of freedom, we have only been able to investigate
small clusters up to four sites\cite{note4sites}.
In such small clusters finite size effects are
obviously relevant. However, we checked that
as far as the criterion for polaron formation is concerned, 
our results are rather insensitive of the boundary conditions
and no qualitative changes occur in passing from three- 
to four-site lattices: In the short range models considered here,
polaron formation is a local, high-energy phenomenon.

The numerical calculations for the Holstein model have
been performed using a slightly modified version of 
the Holstein Hamiltonian (\ref{holham}), in which 
the phonon displacement operator is coupled to the
local electron density fluctuations\cite{rt}:
\begin{equation}
{\cal H}_{e-ph} = g\sum_i (n_i - \langle n_i\rangle)(a_i + a^{\dagger}_i),
\end{equation}
where $n_i = c^{\dagger}_ic_i$ is the number operator for electrons
on site $i$ and $\langle n_i\rangle$ is the mean value of the same
quantity.
This choice removes the trivial coupling between
the mean electron density and the zero momentum phonon mode,
and thus allows for better convergence as far 
as the number of phonons is concerned, without affecting
physically relevant quantities.

\begin{figure}
{\hbox{\psfig{figure=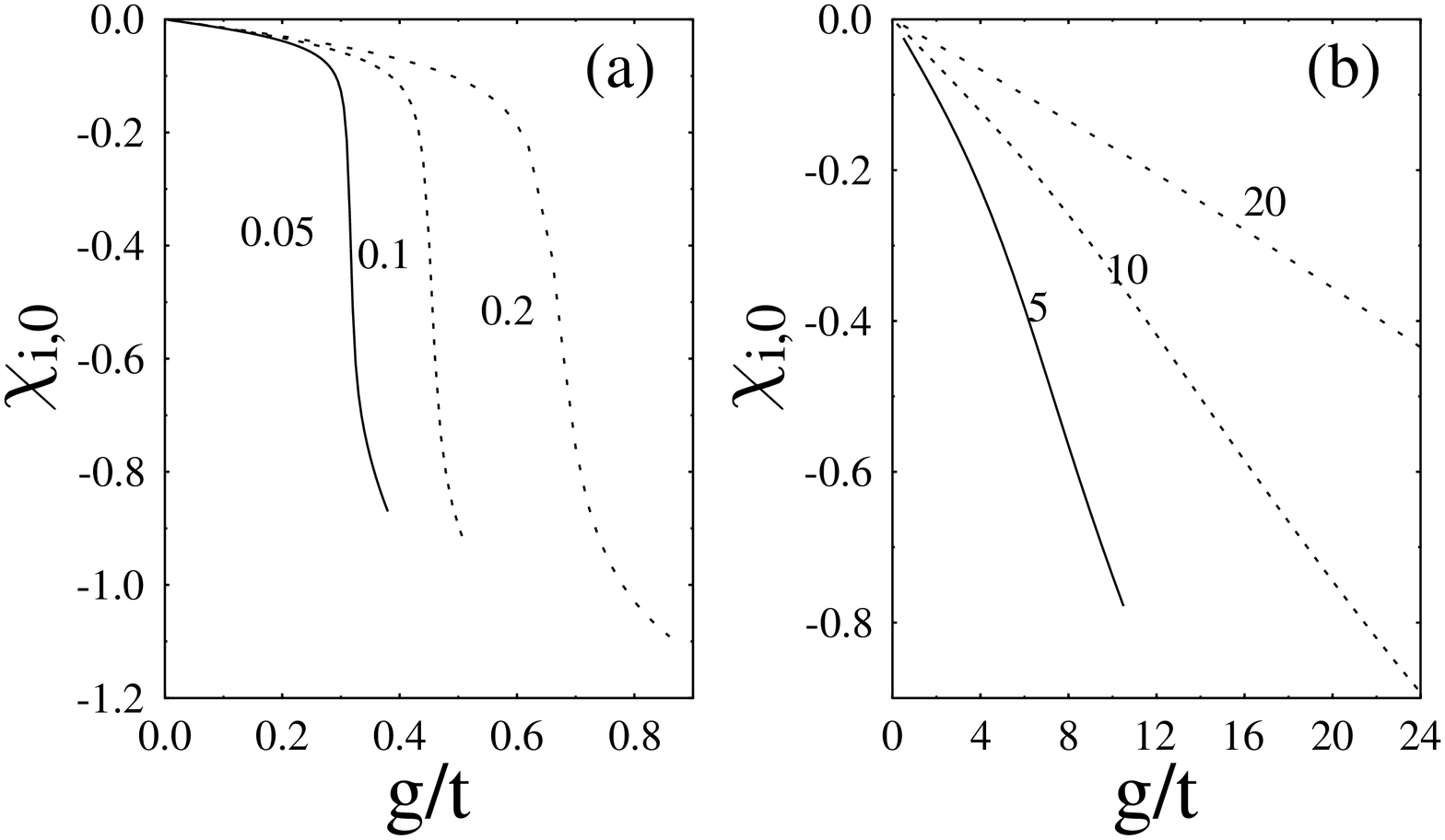,width=5.5cm,angle=-90}}}
\label{Hchi}
\end{figure}
{ \small  FIG. 2: Local-density--displacement correlation
function for the Holstein model and
one electron in a four-site lattice with
periodic boundary conditions.  Lines are labelled by
$\omega_0/t$.}

To extract information on the values of $g$ at
which the polaron crossover begins we analyzed the correlation
function between the electronic density on a site $i$ and the
ionic displacement on the site $i+\delta$
\begin{equation}
\chi_{i,\delta} \equiv \langle \phi_0 \vert
\sum_\sigma c^\dagger_{i\sigma}c_{i\sigma} 
\left( a_{i+\delta}+a^\dagger_{i+\delta} \right)
\vert \phi_0 \rangle
\end{equation}
where $\vert\phi_0\rangle$ is the ground state of the system.

For the Holstein model we report in Fig. 2 the 
behavior of the local-density--displacement
correlation function $\chi_{i,\delta =0}$ for different
values of the adiabatic parameter $\omega_0/t$ as a function
of the $e$-$ph$ coupling constant.
The calculation was performed on a four-site lattice with
periodic boundary conditions.
As is well known, all physical
quantities are smooth functions of the $e$-$ph$ coupling. 
Nevertheless it is evident that the adiabatic regime  
is characterized by a rather sharp crossover, whereas
the increase of $\chi_{i,\delta =0}$ is much slower when 
$\omega_0/t>1$. By calculating non-local correlation
functions with $\delta \ne 0$, we also checked
that, in the Holstein model in one dimension, 
apart from the crossover region, polarons
are always small. In fact the increase of the local
$\chi_{i,0}$ is always accompanied 
in the strong-coupling regime by the decrease of
the non-local correlation functions, showing that
the polaron is so narrow that the presence of a fermion on 
a site is uncorrelated with the ionic displacements on 
neighboring sites.

Since polaron formation is
a crossover without symmetry changes between two phases,
some arbitrariness is unavoidable in defining
a criterion separating the free-electron and the
polaronic regimes. 
In particular we choose the critical $g$ from the point 
of maximum slope of the local-density--displacement correlation
function $\chi_{i,0}$. We checked that different criteria 
(like, e.g., the maximum of the nearest-neighbor-density--displacement 
correlation function $\chi_{i,1}$) provide the same qualitative results.

\begin{figure}
{\hbox{\psfig{figure=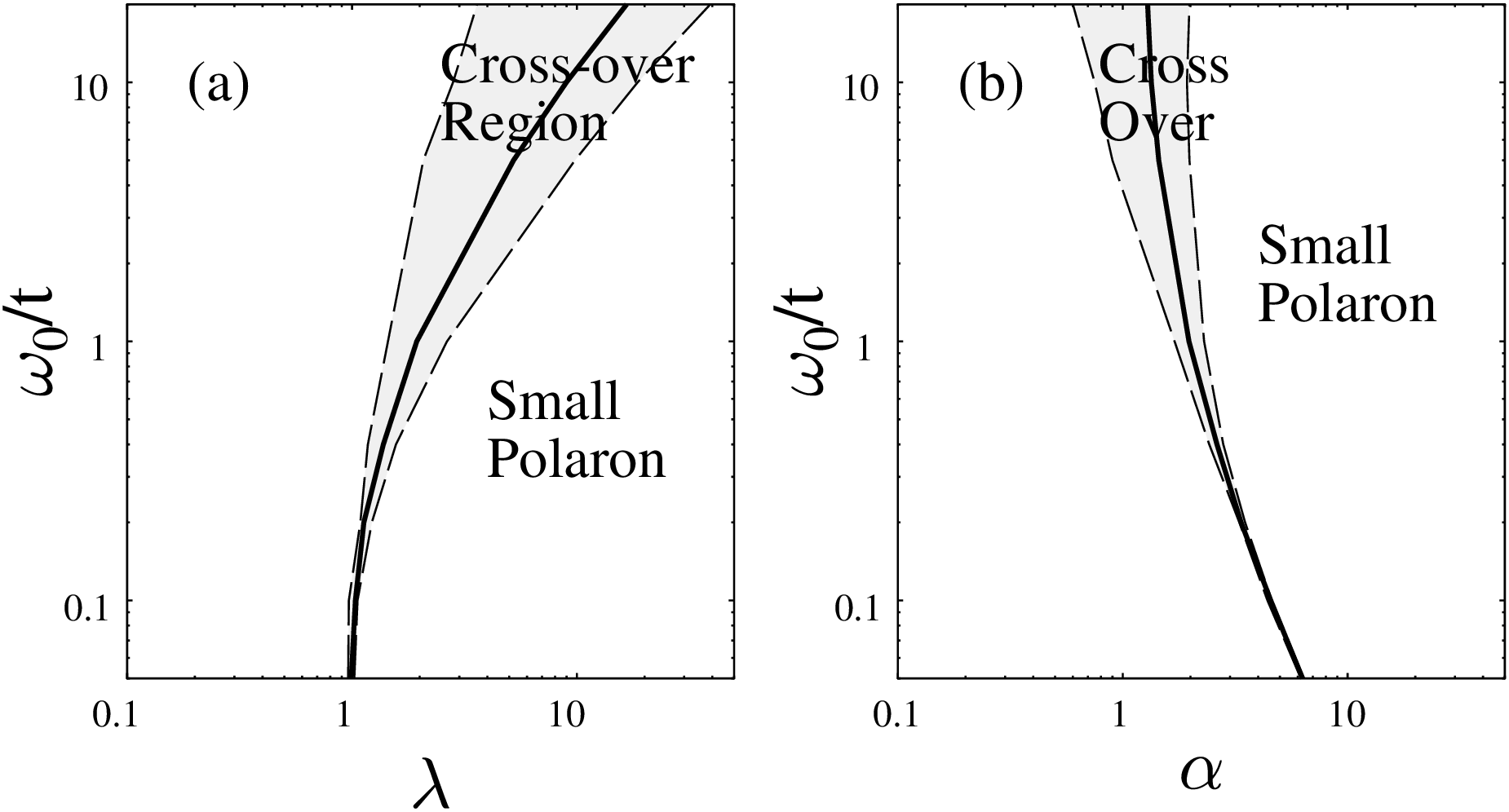,width=5.5cm,angle=-90}}}
{\small FIG. 3: Phase diagram for the Holstein-model
with one electron in a four-site lattice with
periodic boundary conditions:
The crossover region is shaded, while the solid line
is the critical value defined in the text as a function 
of $\lambda$ (a)
and of $\alpha$ (b).}
\label{phdiag}
\end{figure}

From the critical values of $g$ we calculated the
corresponding values of $\lambda$ obtaining the
phase diagram of Fig. 3(a).  
The critical $\lambda$ as a function of $\omega_0/t$ is
indicated with a solid line. We also show the crossover
region, defined as the range of parameters for which 
$\chi_{i,0}$ significantly changes (we calculated the 
second derivative of $\chi_{i,0}$ to numerically 
evidentiate this region).

This clearly shows that
at small phonon frequencies the criterion for having a
polaronic regime is $\lambda=\lambda_c \approx 1$,
whereas at larger phonon frequencies one obtains
$\omega_0/t \approx 2\lambda_c$, which implies 
$\alpha_c \approx 1$.
In Fig. 3(b) the same information of Fig. 3(a) is translated in
terms of $\alpha$. In this way it is made clear that $\alpha_c$ 
is constant in the anti-adiabatic regime $\omega_0 > t$.
Therefore, the exact numerical solution on a small cluster 
confirms the intuitive arguments as well as the perturbative
calculation\cite{notad12}.

The same analysis has been carried out for the SSH model.
While the general scheme is the same, some additional care
has to be used here to check that the adopted values of $g$
do not invert the sign of the hopping parameter or of the
kinetic energy.
In the SSH model, an electron tends to shrink a bond
increasing the effective hopping between two sites. At the 
same time the neighboring bonds are stretched and 
the hopping between the two occupied sites and the
surrounding ones is reduced resulting in a tendency
towards localization. Due to translational invariance
this hopping reduction is translated into an effective
reduction of the quasiparticle bandwidth (i.e. an 
enhancement of the effective mass). 
Eventually the hopping between
the two sites and the rest of the lattice vanishes and
may even change sign. This pathological situation is a
well known feature of the SSH model, which in real systems
never occurs due to higher order corrections to the
expansion of the hopping parameter $t$
in terms of the ionic displacement.
 For all couplings where we find
polaron formation in the SSH model, we checked that 
these pathologies do not occur. We also notice that,
by increasing the phonon frequency, the effective hopping
is relatively less affected, so that larger 
values of $\lambda$ can be reached before the 
zero-hopping pathology is found. As a consequence,
the region with substantial polaronic character
is enlarged.

The results for the
local density-ionic displacement correlation functions do not
qualitatively differ from those of the Holstein model. 
Again one finds that for finite phonon frequencies 
physical quantities are smooth functions of the 
$e$-$ph$ coupling, and that the polaron formation
is a rather fast crossover when
$\omega_0<t$, whereas the crossover is slower for larger
$\omega_0$. In the present case the phase diagram is reported
in Fig. 4. 

This diagram is evaluated making use of the nearest-neighbour
correlation function, which is an increasing function of $g$ in
weak-coupling and it is decreasing in strong-coupling.
The maximum of this function will then be the critical value
for polaron formation.
The crossover region is estimated by the ``width'' of the
same function, whereas the pathological region of
parameters is associated with a negative value of the same 
quantity, as a consequence
of the unphysical negative value of the effective hopping
matrix element.  

As expected from the perturbative
calculation, we find that the polaronic regime is determined
by the condition $\lambda_c =$const. both in the 
adiabatic and the anti-adiabatic regimes\cite{notalambda}.
The specific mechanism of hopping
reduction giving rise to localization in the
SSH model, accounts for this difference with respect to
the Holstein model.

\begin{figure}
 {\hbox{\psfig{figure=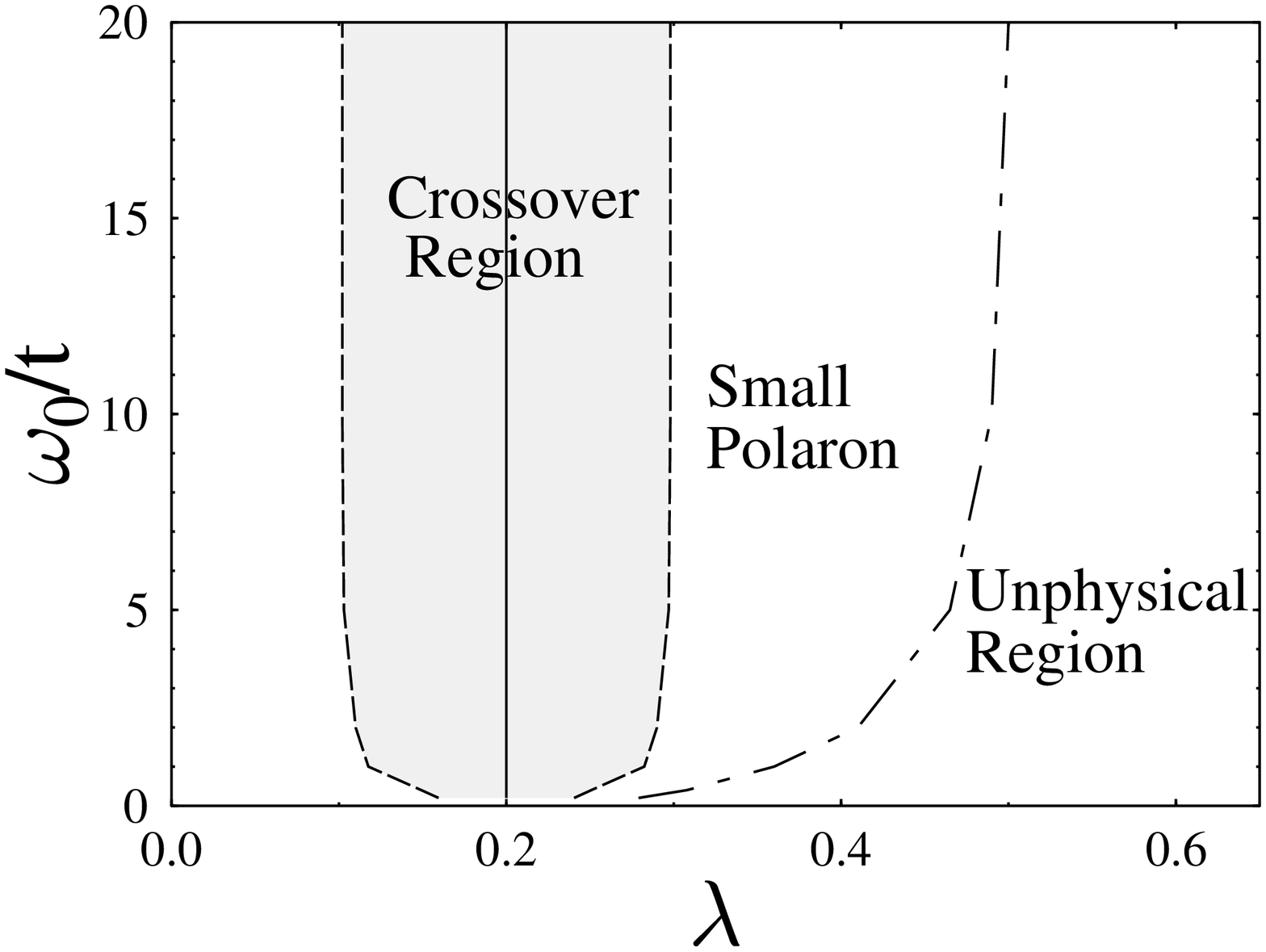,width=5.5cm,angle=-90}}}
{\small FIG.  4: Phase diagram for the SSH-model
with one electron in a four-site lattice with
open boundary conditions:
The solid line is the critical value of $\lambda$ for the
polaronic crossover;
The crossover region is shaded and its boundary is indicated by
the dashed lines. The dot-dashed line is the boundary of the 
pathological region.}
\label{phdiagssh}
\end{figure}

For clarity we begin this discussion by considering the 
fully adiabatic limit\cite{notaadiab}.
In this case, at first order the hopping is reduced by the 
stretching of the bonds by $t^*=t-gu$,
where $u$ is the (dimensionless) value of the bond length
variation in units of $(2M\omega_0)^{-{1\over 2}}$
(notice that for $\omega_0 \to 0$, $g$ has to vanish but
$gu$ stays finite)\cite{notaadiab}.
Since it is possible to show that also
in the SSH model the lattice displacement 
[in units of $(2M\omega_0)^{-{1\over 2}}$]
is proportional to $g/\omega_0$, one finds
\begin{equation}
\label{16}
t^*=t-gu=t\biggl(1-\gamma {g^2\over 2\omega_0 t}\biggr)=t(1-\gamma \lambda).
\end{equation}

In the strong coupling regime the electron is localized on
a single bond and the many-site model becomes equivalent to 
a two-site cluster. Then, for such a small system, an 
analytic solution is possible giving $\gamma =4$
for the constant appearing in Eq.(\ref{16}). 
Thus $\lambda$ determines the reduction of the effective hopping
when the lattice displacement is sizable. 
 
As for the Holstein model case, another
condition needs to be satisfied in order to have a polaronic 
regime: The energy gain due to the polaron formation, i.e. the
typical energy of a self-trapped carrier has to be larger than
the loss in bare kinetic energy associated with the self trapping.
Since in the strong coupling regime the electron localizes on a
single bond we can limit ourselves to a two site-cluster in 
order to evaluate the polaronic binding energy.
The solution of the two-site cluster shows that the polaron 
energy for the SSH model still contains a contribution
from the free electron hopping, arising from the delocalization
of the electron between the two sites of the bond. Then
the ground state energy is given by
$E_0 = -t - 2g^2/\omega_0$. We divide this energy by the free
electron energy $-2t$ to obtain the ratio of the
energy gain associated with polaron formation to the energy loss
associated with the decrease of electronic mobility\cite{notalamhol}.
If we explicitly evaluate the range of $\lambda$ values for which 
\begin{equation}
(-t-2g^2/\omega_0)/(-2t) > 1, 
\end{equation}
we readily obtain $\lambda > 0.25$.
This value coincides with the value at which the hopping matrix
element vanishes according to Eq.(\ref{16}) and to the adiabatic
limit of the parameter $\gamma =4$.
This implies that when $\omega_0 \rightarrow 0$,
the system will have no energetic advantage in localizing the
electron on a bond, unless the pathological condition
$t^*=0$ is reached. According to the physical idea that 
{\it both} a sizable lattice displacement {\it and} an energy gain
from deformation larger than the kinetic energy loss are required
to realize a polaronic state, one should not expect polarons
in the adiabatic limit of the SSH model. Indeed, we carried out the
exact diagonalization of large clusters (100 sites)
in the extreme adiabatic limit finding that the SSH model
does not present any marked polaronic behavior for couplings
smaller than the ``pathological'' $g$'s at which the hopping
changes sign.

For finite phonon frequencies  this picture is modified by
the lattice dynamics. The numerical study shows that
the ground state energy is not strongly effected by
the lattice dynamics: regardless of the value of $\omega_0/t$, 
$\lambda$ larger than $0.25$ remains 
the condition to obtain an energetic
advantage from localization. On the other hand
the effective hopping matrix element
is less severely reduced by the coupling to the lattice fluctuations
and the value of $\lambda$ for which the effective hopping 
becomes zero increases with $\omega_0/t$
(from Fig. 4, one sees that $\gamma \approx 2$ for $\omega_0/t \approx 20$)
Therefore, for finite $\omega_0$, it is
always  possible to find a regime
where the lattice deformation becomes energetically favorable
and a substantial lattice displacement (i.e. hopping reduction)
is present
without having a non-physical vanishing of the hopping. In this region,
which is larger for large phonon frequencies (see Fig. 4),
the electron has a polaronic character for values of $\lambda$
larger than $\lambda_c \approx 0.2$. Finite size effects easily
account for the small quantitative discrepancy between this
value and fully adiabatic estimate $\lambda_c \approx 0.25$.

The above  result shows that $\lambda$ determines both the reduction 
of the hopping integral associated with the lattice distortion 
and the tendency towards localization driven by the energetic
advantage in deforming the lattice.
It is then natural to consider $\lambda$ as the relevant parameter 
for polaron formation regardless of the value of the
adiabatic ratio $\omega_0/t$.
Notice that this finding was also suggested by the perturbative result 
(\ref{msshaad}), showing that
the main corrections to the effective mass are
proportional to $\lambda$ both in the adiabatic and
in the anti-adiabatic regime.

\section{Conclusions}
In the present paper we addressed the issue of polaron formation
in lattice models with extreme short-range $e$-$ph$ interactions,
the Holstein and the SSH models. Our work was devoted to 
settle in a definitive way the case of a single polaron in 
a dynamical lattice. Despite
comprehensive and repeated investigations, this topic
still lacked a clear systematic conclusion,
leading to sometimes contradictory and
confusing statements being reported in the literature.
In particular we clarified, both from analytic qualitative
arguments and from numerical exact calculations, that
$\lambda$ or $\alpha$ are {\it not} by themselves independent
parameters which determine the free-electron or the
polaronic regimes in the $e$-$ph$ models. Indeed we showed
for the Holstein model that both conditions, $\lambda >
\lambda_c \approx 1$ and $\alpha >\alpha_c \approx 1$ 
have to be satisfied in order to realize both
the kinetic energy reduction and the sizable lattice displacement
which characterize the polaronic state. Depending on the
adiabatic ratio $\omega_0/t$, the condition for the polaronic
regime is determined by 
$\lambda > \lambda_c $ when $\omega_0 <t$ and by 
$\alpha >\alpha_c$ when $\omega_0 <t$. 

Comparing our findings with the results of a dynamical 
mean field theory calculation, which is exact in the limit 
of infinite connectivity\cite{ciuchi}, we find 
substantial agreement as far as the value of the parameters
ruling the single-polaron formation in the Holstein model 
is concerned. This clearly indicates that the same physical
picture  extracted here from the numerical calculation
in small (one-dimensional) clusters holds for infinite systems
in higher dimensions as well. This ``universal'' behavior
is a natural consequence of the local character of the small
polarons in the Holstein model.

On the other hand, perturbative and numerical calculations
 for the SSH model lead to the
condition $\lambda > \lambda_c $ irrespective of the
adiabatic ratio. This result, apparently contrasting with
the corresponding criterion in the Holstein model, may be
very naturally understood in terms of a
physical argument based on the kinetic energy reduction 
and the ionic deformation.  Therefore, 
although the final criteria 
are different, the same physical picture
underlies the formation of a single polaron in the two models.

\acknowledgments 
We acknowledge fruitful discussions with Prof. C. Castellani,
Dr. S. Ciuchi, 
and Prof. C. Di Castro. This work was partly supported by the
Istituto Nazionale di Fisica della Materia (INFM).

\end{multicols}


\begin{references}
\bibitem{1dlp} The statement is valid for $d>1$, as it is known that in
$d=1$ for the Holstein model in the extreme adiabatic limit 
a polaronic state is formed for arbitrarily small $e$-$ph$ coupling
constant, and a crossover between a large polaron state and a small polaron
state occurs for $\lambda$ of order one (see Ref. \onlinecite{kabanov}).
\bibitem{notaantiad} The trivial result $m^*=m$ is found
if one trivially takes the $\omega_0 \to \infty$ limit, showing
that an infinitely rigid lattice cannot dress the bare
electronic mass for any finite (and in particular small)
$e$-$ph$ coupling. Physical effects only arise by taking
a large but finite $\omega_0 /t$ ratio.
\bibitem {notaadiab} In order to avoid trivial limiting procedures,
 the adiabatic limit $\omega_0 \to 0$ needs a clear specification. We take
the spring constant $k=M\omega_0^2$ finite, while letting the
ionic mass $M \to \infty$. To keep finite $\lambda = g^2/(2t \omega_0)$,
$g$ has to vanish as $\sqrt{\omega_0}$ giving a divergent $\alpha$.
This leads to a finite value of the lattice displacement
$u=2g/(\omega_0 \sqrt{2M\omega_0})=g\sqrt{2M\omega_0}/k$.
\bibitem{perturb} This argument is only indicative, because
once the mass correction grows large the perturbative result
is no longer quantitatively reliable.
\bibitem{note4sites} A single four-site plaquette
is a one-dimensional structure.  
Besides simplicity, this 
is an additional reason to limit our
analytic treatment to the $d=1$ case.
\bibitem{notad12} We emphasize again that, although in the fully
adiabatic limit in $d=1$ the Holstein model displays a crossover
between a localized large polaron and a localized small polaron 
regime, in our finite frequency calculation on small clusters, 
we cannot distinguish between nearly free electrons and localized 
large polarons. This distinction, however, seems to us irrelevant
in so far we are numerically determining the crossover between
large light carriers and small heavy carriers.
\bibitem{notalambda}Notice
that the numerical value of $\lambda_c \approx 0.2$ 
is different from the result obtained in the adiabatic regime
for the Holstein model $\lambda_c \approx 1$. This is not
problematic because the coupling constants in the two models
have completely different origins and no relation between
their numerical values has to be expected.
\bibitem{notalamhol} This ratio has the same physical meaning 
of $\lambda$ in the Holstein model.
\end{references}
\end{document}